This manuscript is currently under review in Nano Letters

# Toward two-dimensional $MoS_2$ electrets

*Eugenio Lunedei,[1†] Andrea Liscio,[2†] Francesco Borgatti[1] Edoardo Chini[1], Pasquale D'Angelo,[3] Fiorenza Esposito,[3,4] Niccolò Borghi,[5] Matteo Mannini,[5] Denis Gentili,[1]* Luca Seravalli,[3] Massimiliano Cavallini[1]**

1- Istituto per lo Studio dei Materiali Nanostrutturati (ISMN) - Consiglio Nazionale delle Ricerche (CNR) Via P. Gobetti 101, 40129 Bologna, Italy.

2- Istituto per la microelettronica e microsistemi (IMM) Consiglio Nazionale delle Ricerche (CNR), Rome unit, Via del fosso del cavaliere 100, 00133, Roma, Italy.

3- Istituto dei Materiali per l'Elettronica ed il Magnetismo (IMEM) - Consiglio Nazionale delle Ricerche (CNR) Parco Area delle Scienze, 37/A, 43124 Parma, Italy.

4- Department of Mathematical, Physical and Computer Sciences, University of Parma, Parco Area delle Scienze 7/a, 43124 Parma, Italy

5- Dipartimento di Chimica " U. Schiff" Università di Firenze, Via della Lastruccia 3-13, 50019 Sesto Fiorentino, Italy.

*** e-mails: massimiliano.cavallini@cnr.it; denis.gentili@cnr.it**

## Abstract

Here, we show that controlled sulfur-vacancy engineering converts monolayer $MoS_2$ into an electret, imparting the ability to store quasi-permanent electrostatic charge within a single atomic layer. Sulfur vacancies are generated with submicrometer spatial control by stamp-assisted electrode-free electrochemical nanolithography, yielding programmable defect densities from $10^{10}$ to $10^{13}$ cm$^{-2}$. The resulting vacancy domains act as deep electron traps and produce surface charge densities up to 1 μC cm$^{-2}$, with charge retention in the order of hundreds days under ambient conditions. Kelvin probe and electric force microscopies directly reveal stable electrostatic patterns that replicate the lithographic motif. The same vacancy landscape simultaneously defines exciton-quenching regions, generating co-localized optical and electrostatic contrast and reducing the apparent exciton lifetime from 15 ns to 180 ps through enhanced nonradiative recombination. The persistence of the optical response over one year identifies sulfur vacancies, rather than transient charge states, as the origin of the patterned functionality. These results establish defect-engineered $MoS_2$ as the first two-dimensional electret and demonstrate that atomic vacancies can be exploited as functional elements for encoding electrostatic and excitonic behavior in a single atomic layer.

## Introduction

Over the past two decades, the discovery of two-dimensional (2D) materials has revolutionised many areas of materials science and condensed matter physics[1]. Reduced dimensionality fundamentally alters the electronic structure by inducing strong quantum confinement, enhancing surface and interface effects, and giving rise to characteristic low-dimensional densities of states[2], thereby enabling several unique functionalities[3-5]. However, these same features render 2D materials exceptionally sensitive to defects, such that even low defect densities dramatically alter their chemical, electronic, and optical properties.

Although defects are traditionally considered detrimental to material performance and stability, some of them offer unexpected opportunities that are utilised as an additional degree of freedom in materials engineering. In particular, atomic vacancies and heteroatoms, which are the most diffuse defects, introduce new functions and physical phenomena that are impossible in pristine material; these include opportunities in ion storage[6], catalysis[7] and quantum capacitance[8], among others. Yet many defect-driven properties have been demonstrated only qualitatively, because their intensity and control are insufficient for practical applications. Therefore, defect engineering is crucial for managing and utilizing these effects in technological and fundamental applications.

Here, we exploit this paradigm to introduce an unexplored functionality by demonstrating that a $MoS_2$ monolayer can host spatially controlled and programmable long-lived charge trapping, thereby conferring electret functionality, i.e., the ability to store quasi-permanent electrostatic charge, on a $MoS_2$ monolayer[9], a property not achieved in any actual 2D system.

Although charge trapping in 2D semiconductors has been observed in various 2D contexts[10-13], the charge densities and retention times achieved so far have been insufficient for electret applications. We overcome these limitations through fine defect engineering using electrode-free electrochemical nanolithography[14], a technique that generates spatially localised charged defects in $MoS_2$. This approach transforms two features typically associated with reliability concerns, namely electrically induced defects[15] and charge trapping[10], into a functional property in 2D $MoS_2$. Samples were characterised by the typical methods used to study defects and electrect: optical and fluorescence microscopy (FM), atomic force microscopy (AFM), Raman, X-ray photoelectron spectroscopy (XPS), Kelvin probe force microscopy (KPFM), electric force microscopy (EFM) and time-resolved fluorescence (TRF).

$MoS_2$ monolayers crystals (2D-$MoS_2$) were grown on Si/$SiO_2$ substrates by chemical vapor deposition[16,17].

To fabricate spatially defined defect domains, we employed electrode-free anodic oxidation nanolithography (EFLAO), a recently developed technique for the oxidative etching of 2D materials [14], which also enables fine control over defect formation. In EFLAO, a conductive probe is brought into contact with the surface in a high-humidity environment (RH>90%), forming a nanoscale electrochemical cell. Unlike conventional electrochemical nanolithography, EFLAO operates through capacitive coupling between the conductive probe and the sample, applying an AC bias[14]. Originally implemented as a scanning probe technique, we adapted EFLAO for parallel processing by replacing the probe with a stamp, thereby enabling centimetre-scale patterning in a single step[18]. Figure 1 shows the scheme of the process.

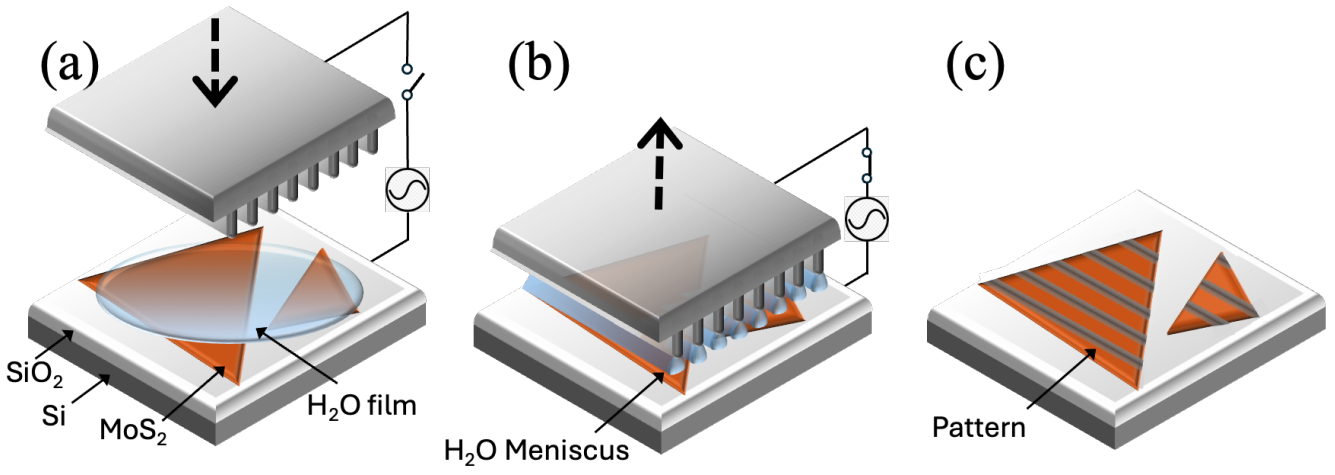

**Figure 1.** Scheme of stamp-assisted electrode-free local anodic oxidation nanolithography. a) A thin water layer is formed in highly humid conditions (Relative Humidity >95%). b) Stamp motifs create a meniscus through capillary force upon contact with the surface, forming an array of nano-electrochemical cells. Applying an AC bias voltage induces electrochemical etching within these nano-cells. c) The reaction occurs exclusively beneath the stamp protrusions, resulting in the patterning of surface structures that replicate the stamp relief features.

The morphological effects of EFLAO treatment on 2D-$MoS_2$ were examined using contact AFM, a setup chosen to suppress electrostatic artefacts, such as those that can occur in intermittent-contact imaging, especially when measuring crystal thickness[19]. For treatment durations shorter than 150 min, no detectable morphological changes were observed on 2D-$MoS_2$ (Figure 2a). No trace of additional material or local reorganisation was observed. At longer treatment durations, a <0.5 nm increase in local roughness was observed (Figure S2), indicating the beginning of lattice deformation. For this reason, only treatments of less than 150 min were considered in subsequent analyses.

X-ray photoelectron spectroscopy (XPS) measurements show that, as is typical for CVD-grown $MoS_2$, the pristine material already contains a fraction of native defects (~5%), identified as sulfur vacancies (Vs)[20]. Despite a small increase in the Vs signal (not experimentally significant, as the variation falls within the experimental error), a comparison between pristine and treated 2D-$MoS_2$ shows no significant changes in either binding energies or relative peak intensities (Figure S3 and Table S1), indicating that any compositional modification induced by the treatment involves less than 1% of the atomic species.Confocal Raman spectroscopy, performed on pristine 2D-$MoS_2$ reveals the two characteristic first-order vibrational modes, namely the in-plane $E^1_{2g}$ mode at ~384 $cm^{-1}$ and the out-of-plane $A_{1g}$ mode at ~405 $cm^{-1}$ (Figure 2c), consistent with the monolayer nature of our samples[21]. After EFLAO treatment, the overall Raman line-shape remains essentially unchanged (Figure 2c), confirming that the crystal lattice is preserved and no significant structural degradation occurs during processing. Nonetheless, treated samples systematically display a bit larger $A_{1g}$ - $E^1_{2g}$ mode separation, primarily arising from a red shift of the $E^1_{2g}$ peak by approximately 2.0 cm$^{-1}$ (Figure 2c inset). This spectral shift is consistent with a mild, fractional desulfurization process, corresponding to an estimated average Vs spacing exceeding 10 nm [21,22].

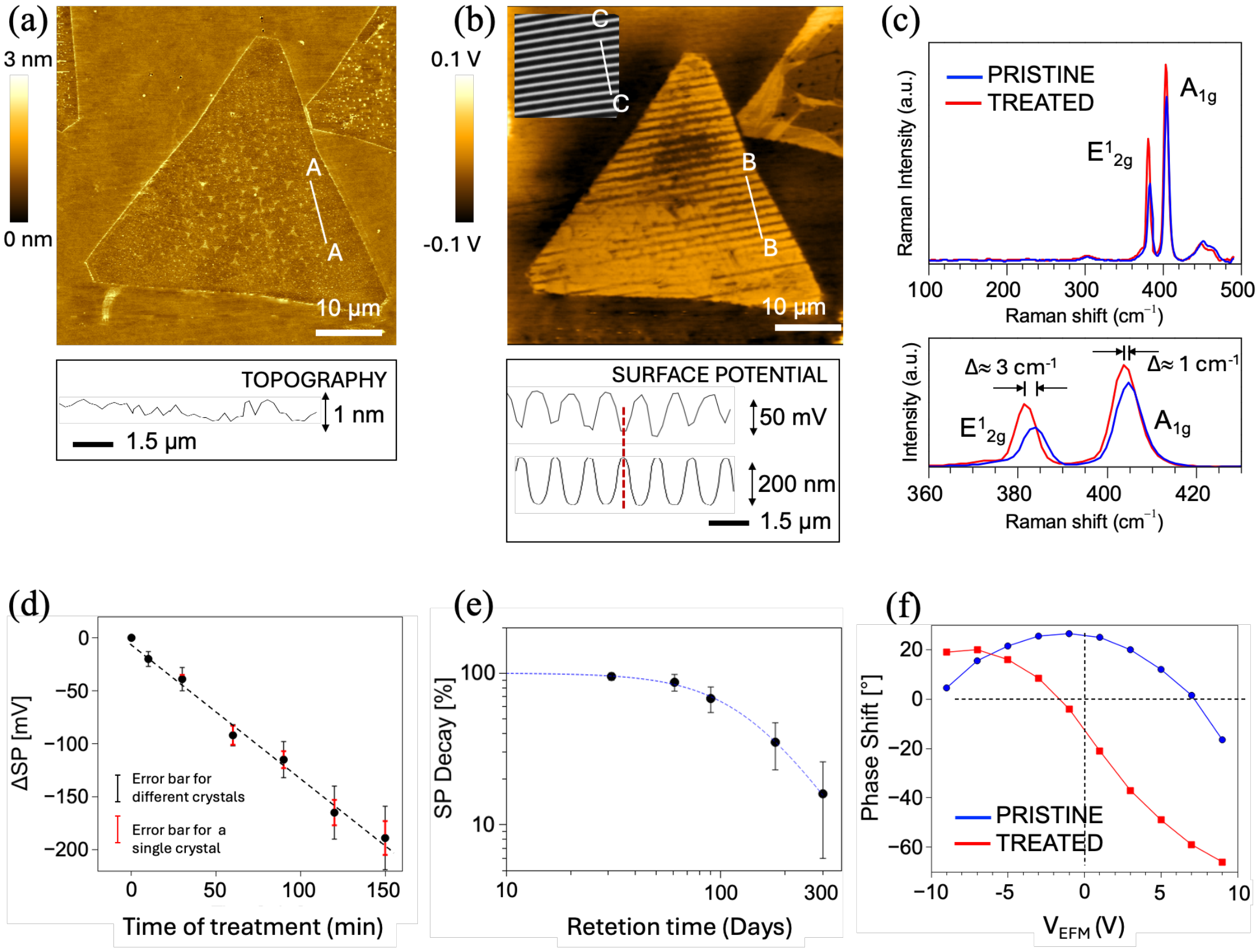

**Figure 2.** a) Contact AFM topographic image of a $MoS_2$ flake, one monolayer thick, after EFLAO treatment of 60 minutes, close to an untreated flake deposited on $Si/SiO_2$ substrate. b) Corresponding surface-potential map obtained by KPFM. The inset shows a morphological image of the stamp. c) Raman spectra (λexc=532 nm) of pristine (blue) and treated (red) monolayer crystals; (inset) Zoom on the $E^1_{2g}$ and $A_{1g}$ bands spectral region. Upon patterning, $E^1_{2g}$ peak is observed to shift from ~383.8 cm-¹ to ~381.7 cm-¹ (Δλ~2.1 cm-¹). d) Evolution of surface potential with respect untreated zones upon treatment with different application times. Black bars refer to the error considering different crystals, red error bars refer to the error inside a single crystal. Dash line is a guide for the eyes. e) Long-term trend (expressed as percentage) of surface potential, maintaining the sample in air at ambient conditions and exposed to light. The blue curve represents the Hill-type fitting with exponent n≈2. f) Electric force microscopy phase–bias curves for pristine (black curve) and treated (red curve) 2D-$MoS_2$.

With treatments longer than 150 min, the defect density reaches a level at which multi-vacancy complexes are likely to form, leading to a lattice reorganisation and consequently a local increase in morphological roughness, as is usually observed in the formation of high-density Vs[5,23] in $MoS_2$ and evidenced by Raman investigation[21,22].

The results of morphological and spectroscopic investigationsaligns with reports on moderate electrochemically assisted desulfurisation of $MoS_2$[23,24], suggesting the early formation of multiatomic defects that compromise the quality of the $MoS_2$ lattice during treatments longer than 150 min.

The electronic effects of EFLAO patterning were investigated using KPFM and EFM.

Unlike morphological (no)effects, even after short treatments, the surface potential (SP) maps obtained by KPFM show a strong modulation (Figure 1b), perfectly replicating the stamp features (Figure 1b). The treated regions form parallel stripes approximately 250 ± 50 nm wide with a 1500 nm pitch, exhibiting a more negative SP than untreated areas. The magnitude of the SP contrast measured within a flake ($\Delta SP = SP_{Treated} - SP_{Untreated}$) increases linearly with treatment duration, from $\Delta SP = -23 \pm 7$ mV after 10 min to $\Delta SP = -180 \pm 30$ mV after 150 min (Figure 2d). At longer durations, the onset of morphological roughening reduces measurement reproducibility. The linear increase in ΔSP aligns with a non-interacting formation of isolated atomic Vs, resulting in a low vacancy density. Within 150 min of treatment, no signs of preferential formation of multiatomic Vs were observed.

The negative polarity of ΔSP indicates the accumulation of trapped negative charge (i.e., electrons) in the patterned areas. The SP steps at the stripe boundaries are relatively sharp, showing strong lateral confinement and minimal diffusion of trapped charge across the monolayer. The polarity and periodicity of the potential contrast are highly reproducible across different samples, with variability within a flake below 10% (red error bars in Figure 2d). Slight differences between flakes (black error bars) arise from local variations in the stamp–surface gap, caused by multilayer regions or minor environmental fluctuations, indicating a strong sensitivity of the process to the experimental conditions and substrate quality.

Using an infinite plane electrostatic approximation[19], the trapped charge density was estimated to range from ~130 ± 10 nC/cm$^2$ after 10 min of treatment to −1000 ± 50 nC/cm$^2$ after 150 min. These charge densities are comparable to or exceed those of many conventional electrets[25,26], yet they remain confined to a single monolayer. Assuming each formed defect traps one electron, the number of sulfur vacancies ($N_{Vs}$) per surface unit increases from 8.1 x $10^{11}$ $N_{Vs}$/cm$^2$ after 10 minutes of treatment to 6.3 x $10^{12}$ $N_{Vs}$/cm$^2$ after 150 minutes, corresponding to a defect formation rate of 3.7 x $10^{10}$ ($N_{Vs}$/cm$^2$)/min.

Under ambient conditions (RH 50–75%) and light exposure, ΔSP gradually decays with ageing. Within 300 days, ΔSP decreases from 100% to ~20%, independently of its initial value (Figure 2e). To phenomenologically describe the discharging process, we fitted the temporal evolution of the normalised surface potential contrast using a Hill-type function $\Delta SP(t)/\Delta SP(t_0)=[1+(t/\tau)^n]^{-1}$, where $\Delta SP(t_0)$ is the initial surface potential contrast, τ is the characteristic relaxation time, and *n* is the Hill exponent, which quantifies deviations from first-order kinetics. The resulting decay is well described by a Hill-type function (Figure 2e) with a characteristic relaxation time of τ =134 ± 4 days and an exponent $n \approx 2$. Such an exponent is usually linked to spatially correlated relaxation dynamics in disordered systems, rather than with independent detrapping from isolated trap states[27].

The electrostatic origin of the patterned contrast was confirmed by electric force microscopy through measuring the phase shift relative to the applied bias (Figure 2f) [28]. While pristine flakes show the typical quadratic dependence of the phase shift on the applied bias, characteristic of purely capacitive interactions (Figure 2f, blue curve), the treated regions show an almost linear dependence (Figure 2f, red curve), indicating electrostatic coupling between the probe and static surface charge. The negative phase shift further confirms the accumulation of excess negative charge in the patterned domains [29].

Collectively, these results demonstrate the efficiency of the EFLAO treatment in precisely tailoring the local electronic properties at a submicrometric scale by creating Vs in the $MoS_2$ lattice, which act as charge-trapping centers.

Because the optical properties of 2D-$MoS_2$ are highly sensitive to the nature and concentration of defects[30], we conducted complementary steady-state fluorescence micro-imaging and time-correlated fluorescence micro-spectroscopy to investigate the optical effects induced by EFLAO on individual 2D-$MoS_2$. All measurements were taken at the center of the crystals to avoid possible edge-related effects, which are known to influence large-area $MoS_2$ monolayers (Figure 3a,b)[21].

In fluorescence micro-spectroscopy, 2D-$MoS_2$ exhibits a broad emission spectrum with a distinct band centred at ~665 nm (Figure 3C), which arises from a single emitting state associated with the A-exciton transition at the K point of the Brillouin zone.

Following EFLAO treatment, fluorescence images reveal an optical pattern of parallel lines arising from fluorescence quenching (dark regions), corresponding to the treated zones identified by KPFM (Figure S5).

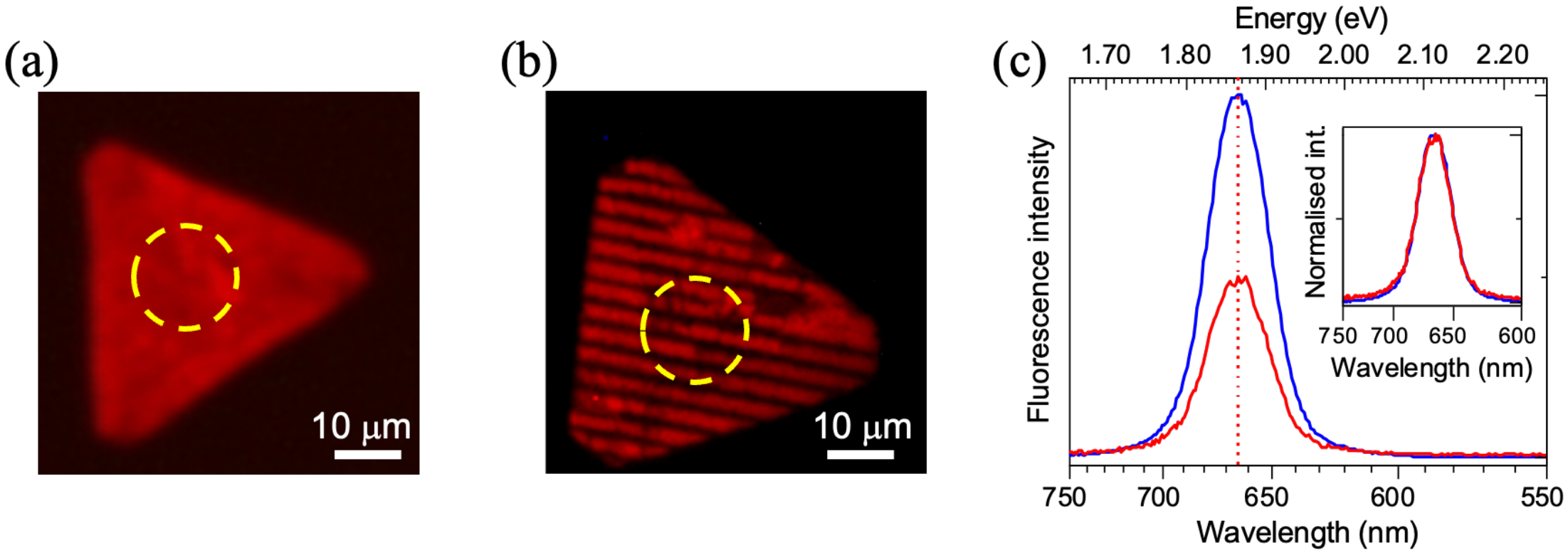

**Figure 3.** Emission micro-spectroscopy. Fluorescence micro-image ($\lambda_{exc}$ = 535 nm, integrated detection spectral range $\lambda_{det}$ > 590 nm) of (a) a pristine and (b) treated monolayer crystal of $MoS_2$. The circle (diameter ~20 μm) indicates the region where the spectra were collected. c) corresponding steady state fluorescence spectra of pristine (blue) and treated (red); (inset) relative intensity normalized plot to evidence the equivalent spectral shape.

Comparative spectral analysis (Figure 3c) shows that the overall emission intensity decreases in the treated samples. Although the treated area identified by KPFM accounts for ~25% of the surface, the fluorescence intensity is reduced by ~50% relative to the initial value, with no alteration in the spectral band shape or peak position (Figure 3c, inset). This behavior indicates that the process induces strong fluorescence quenching, mainly confined to the treated regions, without affecting the character of exciton radiative relaxation in the surrounding untreated areas, which originates from a single emitting state[30]. Additionally, in the fluorescence images, the dark stripes usually appear broader than in KPFM. We investigated exciton dynamics and recombination pathways in 2D-$MoS_2$ using time-resolved fluorescence.

The temporal evolution of the fluorescence spectrum of a pristine monolayer, following a sub-100ps excitation pulse ($\lambda_{exc}$=402 nm), shows a unique emission band centred at ~665 nm. The emission of the pristine crystal decays with a relatively long time-constant (Figure 4a, observed range 0–10 ns) while both the spectral shape and the peak position remain unchanged even on a longer time-window (Figure 4a, intensity map, observed range 0-37 ns). Fluorescence transients recorded at the emission maximum give an intensity-averaged decay time of 10–15 ns, consistent with reported values for high-quality $MoS_2$ monolayer[30] (Figure 4c). However, even in pristine crystals, we observed a multiexponential recombination behaviour, indicating the presence of a pre-existing distribution of surface defects that affects the local value of $MoS_2$ exciton diffusion length (see Figure S6 for decay fitting details).

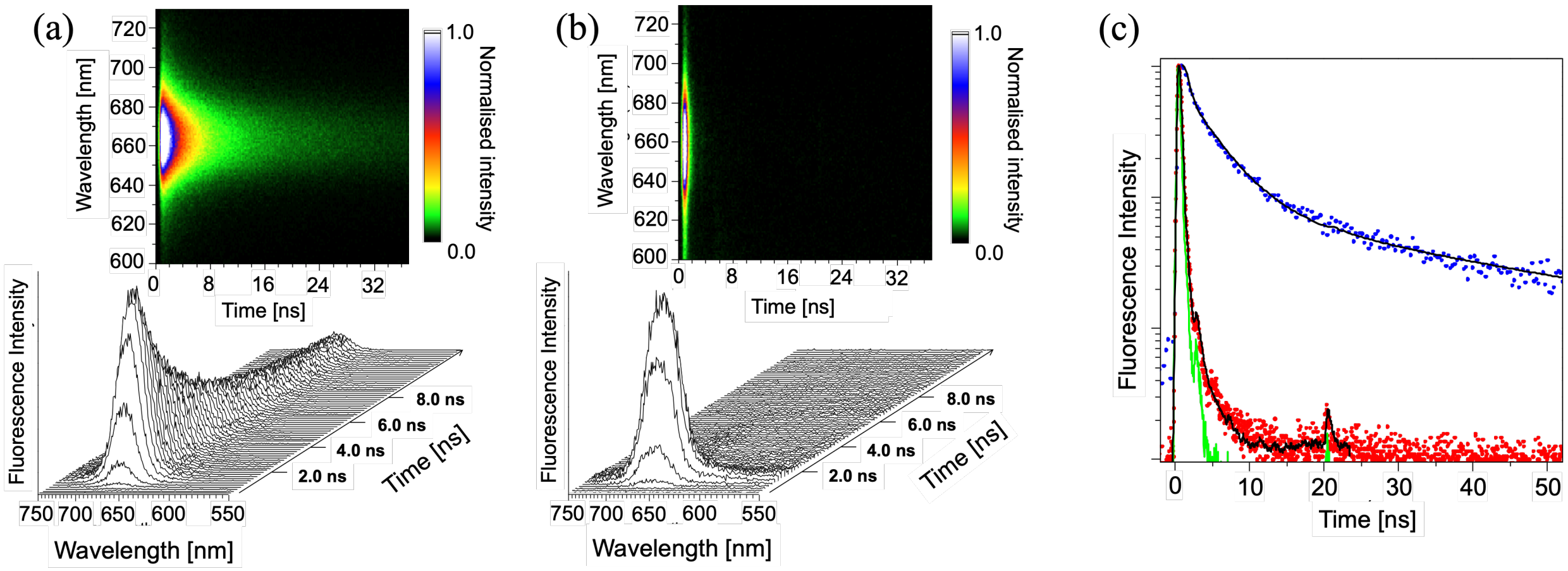

**Figure 4.** Time-resolved fluorescence spectroscopy. a) Sub-ns time evolution of fluorescence spectrum of *pristine* 2D-$MoS_2$ under microscope (objective 100x, $\lambda_{exc}$=402 nm, f=5 MHz, time separation between spectra: $\Delta t$ = 444 ps). False color normalized intensity map, traced up to 37 ns ($\Delta t$ = 74 ps) b) Same visualization for *treated* sample. c) Normalized micro-fluorescence decays both of pristine (blue) and treated (red) monolayer crystal of $MoS_2$ ($\lambda_{exc}$=402 nm, f=4 MHz, $\lambda_{det}$=665 nm, $\Delta t$ = 74 ps), instrumental response function (IRF, green) and relative best fit exponential convoluted curve (black line).

Treated crystals behave differently: their decay becomes much faster, with characteristic times reduced to ~180 ps, while the emission spectrum itself remains unchanged (Figure 3b, and FigureS7 for details). In addition to a reduction in overall intensity, a small shoulder was observed at ~620 nm (~2.0 eV), attributable to B exciton emission [24,31]. This behaviour is consistent with the introduction of defect-assisted non-radiative recombination channels within the printed regions[32], which is associated with a high density of Vs.[30]

The apparent widening of stripes in fluorescence images and the pronounced reduction in the intensity-averaged decay-time in the patterned samples can be qualitatively explained by considering the concurrent effects of defect distribution produced by the process and the natural exciton diffusion length. Because of a locally high defect density inside the printed regions, excitons are expected to decay entirely through non-radiative pathways, creating optically non-emissive “dark zones”.

Under these conditions, the measured fluorescence should come exclusively from the unmodified regions, i.e. the “bright stripes”, which should retain the decay characteristics of the pristine material. Since the width of the bright stripes (~1250 ± 50 nm) is comparable to (or slightly shorter than) the exciton diffusion length in 2D-$MoS_2$ (1000-1500 nm[33]), excitons formed within the bright stripes recombine mainly radiatively. However, owing to their long pristine diffusion length, they can reach defect-rich regions, thereby increasing the effective non-radiative deactivation rate. Indeed, this reduces the observed decay time. This effect increases progressively for excitons formed near the printed stripes. Excitons created near the defect-rich boundaries are almost completely quenched. As a result, dark regions appear wider in fluorescence imaging than the size of the printed zones, and the decay curve of the treated sample becomes significantly faster than that of the pristine material. The resulting fluorescence decay represents a continuous distribution of lifetimes, reflecting excitons generated at different distances from the defect-rich boundaries and subject to progressively stronger non-radiative losses.

Notably, unlike the electric response, the suppression of optical emission in the printed area and the shortening of the decay time in the untreated area remain stable over time and show no evidence of ageing, even after one year of shelf time. This long-term stability indicates that the quenching sites formed by EFLAO are not governed by variable charge states, supporting their assignment to Vs. This behavior shows that nanoscale patterning can be used to manipulate non-local exciton dynamics by shaping the spatial distribution of defects. Combined SPM and spectroscopic analyses show that EFLAO treatment generates lattice defects that act as highly efficient trap centers, producing a robust and long-lived charge-trapping phenomenon in the treated $MoS_2$ crystals and a permanent local quenching of fluorescence. Both experimental observations and thermodynamic considerations converge on identifying desulfurization as the dominant reaction pathway, as Vs are the most readily formed defects[4] and the process proceeds without changing the oxidation states of Molybdenum or Sulfur[23]. The EFLAO process selectively forms sulfur vacancies at the stamp-defined features; these vacancies act as deep electron traps[34] while simultaneously quenching the photoluminescence of the 2D-$MoS_2$[30]. Within this framework, Sulfur removal involves a coupled proton–electron transfer at surface sulfur atoms (S*), producing gaseous $H_2S$ and leaving behind an atomic sulfur vacancy [23] (Scheme S1). We hypothesize that the extreme local conditions of EFLAO activate water oxidation, thereby providing the protons and electrons necessary to sustain the desulfurization reaction and the successive electron trapping. This mechanism is consistent with reports of electrochemically assisted desulfurization in both conventional[23] and stamp-mediated configurations[24] and explains the Vs charging. However, because EFLAO cannot be treated as a conventional electrochemical system and given the extreme local conditions, we do not exclude the possibility of other mechanisms.

In conclusion, we establish defect-engineered monolayer $MoS_2$ as the first two-dimensional electret. Sulfur vacancies generated with sub micrometer precision create a stable population of deep electron traps capable of storing electrostatic charge for months under ambient conditions, while simultaneously defining exciton-quenching landscapes. The coexistence of programmable electrostatic and optical functionality within a single atomic layer demonstrates that atomic vacancies can serve not only as electronic defects but also as design elements for device operation. These findings elevate defect engineering from a means of property tuning to a strategy for creating emergent functionality in two-dimensional semiconductors.

**Materials and Methods:** See supplementary Material

**Supporting Information**. XPS spectra; relative percentage of the $MoS_2$ species; Optical characterization of $MoS_2$ monolayer crystals; Topography of 2D$MoS_2$ crystals treated for time; Direct comparison from surface potential images and fluorescence microscopy image of patterned samples; Pristine and treated monolayer crystal, fluorescence decay fitting parameters and spectra.

AUTHOR INFORMATION

**Corresponding Author**

* e-mails: massimiliano.cavallini@cnr.it; denis.gentili@cnr.it

**Author Contributions**

The manuscript was written through contributions of all authors. All authors have given approval to the final version of the manuscript. E.L. and A.L. contributed equally.

**Funding Sources**

European Union– Next Generation EU from the Italian Ministry of University and Research. Project PRIN 2022SRHPF2 Titled Molecular Assisted atom Vacancies Arrangement to modulate Magnetism in 2D transition metal dichalcogenides″ (MAVAM). L.S. acknowledges the support of the European Union, Next Generation EU, Mission 4, Component 1, through the MUR PRIN2022 project “2DIntegratE” (2022RHRZN2).

**ACKNOWLEDGMENT**

We thank Alessandro Surpi for the assistance in the Raman measurements, Giovanni Attolini and Matteo Bosi for their work in the CVD growth of $MoS_2$.

# Toward two-dimensional $MoS_2$ electrets

*Eugenio Lunedei, Andrea Liscio, Francesco Borgatti, Edoardo Chini, Pasquale D'Angelo, Fiorenza Esposito, Niccolò Borghi, Matteo Mannini, Denis Gentili*, Luca Seravalli, Massimiliano Cavallini**

## SUPPLEMENTARY MATERIAL

**Experimental Section**

***Growth $MoS_2$ Monolayer Crystals:*** $MoS_2$ samples were prepared by CVD assisted by molybdenum liquid precursors on a $SiO_2$(280nm)/Si substrates at 820°C, using a spin-coated solution of $(NH_4)Mo_7O_{24}$, NaOH and Iodixanol. The substrates with the spun solution were inserted into a quartz tube and exposed to sulphur atoms carried by a nitrogen flow[17,35]. The samples consisted of triangular flakes with lateral dimensions of 10–100 μm and a thickness of 0.9 ± 0.2 nm. In some cases, additional features were observed atop the larger monolayer domains, including small crystallites, micron-scale monolayer islands 1–2 nm thick, and, rarely (<1%), multilayer pyramidal structures. Figure S1 shows typical optical micrographs recorded in bright field, dark field, and fluorescence microscopy of the $MoS_2$ flakes. Fluorescence imaging reveals the quenching of PL in correspondence with crystallites and pyramidal structures (FigureFigureS1c). In Multilayer structures, the quenching is consistent with the transition in $MoS_2$ electronic structure from a direct bandgap in monolayer to an indirect bandgap in multilayers[36], resulting in reduced radiative recombination efficiency.

***Parallel electrode-free anodic oxidation nanolithography****: 2D-$MoS_2$* Patterning were performed using electrode-free anodic oxidation nanolithography (EFLAO)[14]. EFLAO relies on high-frequency alternating current operation (>10 kHz) and capacitive coupling, such that only a conductive stamp is required. The method was implemented in a parallel configuration[37,38]. The experiments were carried out using a custom-built setup identical to that employed for parallel electrochemical nanolithography, as described in detail elsewhere [18]. Relative humidity was continuously monitored using a commercial hygrometer (Thermopro). Patterning was achieved by bringing the conductive stamp into conformal contact with the sample surface and applying an alternating bias of 15 V at 50 kHz for a defined duration under high-humidity conditions (95% relative humidity). A schematic representation of the process is provided in Figure 1 of the text. The quality and morphological uniformity of the substrate were found to be critical parameters. In particular, the presence of surface outgrowths or locally thick features markedly reduced both the efficiency and reproducibility of the oxidation process.

***Confocal Raman spectroscopy (μ-Raman):*** Micro-Raman spectra were acquired in backscattering configuration using a Renishaw InVia/1000 Raman microscope equipped with a solid-state laser operating at a wavelength of 532 nm. The excitation beam was focused onto the sample surface through a 100× objective lens, yielding a lateral spatial resolution of approximately 1 μm. All measurements were performed under ambient conditions. To minimize laser-induced thermal effects and avoid unintentional modification of the atomically thin layers, the incident laser power at the sample surface was limited to 80 μW. Spectra were collected at selected points and, where needed, across extended areas of the flakes to assess spatial homogeneity. The spectral calibration of the system was routinely verified using a reference silicon substrate before data acquisition.

***Stamp fabrication and characteristics.*** The stamp employed for the nanolithography experiments was obtained directly from the metallic reflective layer of commercially available blank compact discs, which was used without further pattern modification. This metallic layer was mechanically supported by a 3 mm-thick film of polydimethylsiloxane (PDMS; Sylgard 184, Dow Corning), providing mechanical compliance and ensuring conformal contact with the sample surface during processing. The stamp surface exhibits a one-dimensional grating morphology originating from the optical data tracks of the compact disc. As illustrated in the inset of Figure 2b, the pattern consists of parallel linear features with a periodicity of approximately 1.5 µm. Individual lines present an apex width of about 200 nm and a groove depth of roughly 220 nm.

*AFM, KPFM, EFM Measurements:* Scanning probe microscopy measurements were performed using a Bruker Multimode 8 system for all atomic force microscopy (AFM), Kelvin probe force microscopy (KPFM), and Electrical force microscopy EFM characterisations. All experiments were conducted in ambient air under laboratory conditions. KPFM measurements were acquired in amplitude-modulated tapping mode using a two-pass lift configuration. During the second pass, the tip was lifted by 20–50 nm relative to the surface. Topographic and electrical images were collected simultaneously using standard acquisition parameters optimised for stable operation on the investigated samples. Post-acquisition processing was applied to compensate for piezo-scanner distortions using the Gwyddion software package. Topographic images presented in the main text were corrected using global plane levelling, whereas KPFM maps were processed using line-by-line plane levelling to improve visual clarity while preserving intrinsic contrast.[39]

***Estimation of trapped charge density***. The trapped charge density (σ) was estimated using a parallel-plate capacitor model[19]: $\Delta SP = \sigma h_{eff}/\varepsilon_0 \varepsilon_r$. Where: ΔSP is the surface potential contrast, σ is the surface charge density, $h_{eff}$ the effective electrostatic thickness (one monolayer ~0.85 nm) and $\varepsilon_r$ is the relative permittivity (~5).

***X-ray Photoelectron Spectroscopy.*** XPS measurements were performed using a non-monochromatic Al $K_\alpha$ X-ray radiation (VSW-TA10; Al $K_\alpha$, $\lambda$ = 1486.6 eV) operated at 144 W (12kV , 12 mA) and a VSW-HA100 hemispherical analyser mounting a 16-channel detector and set with a pass energy of 22eV; all the spectra were analysed using the software CasaXPS 2.3 and calibrated with respect to the adventitious carbon in the C 1*s* core level region and fixed at 248.8 eV. The spectral deconvolution analysis of Mo 3*d* signals was carried out using an asymmetric Lorentzian line shape, while for the S 2*s* and S 2*p,* a combination of a Gaussian and Lorentzian line shape was employed instead.

***Steady State Fluorescence Microspectroscopy.*** Fluorescence spectra of $MoS_2$ crystals, as well as theirfluorescence images, were recorded by means of a modified Nikon Eclipse80i epifluorescence microscope. The standard trinocular turret of the microscope was customized in order to mount a home-made optical system (composed of a 50 mm focal quartz lens focused into a UV-Vis/NiR optical fibre (200-2200 nm), 550 µm core, Thorlabs) to feed the fluorescence signal into an Avantes AvaSpec-2048 CCD spectrometer (2048 pixels array, DLC UV/Vis, 200-1100 nm range, 10 µm or 100 µm entrance slit, software programmed. This allowed to record fluorescence images and collect the photoluminescence signal from $MoS_2$ crystals under a 20x-100x objective. For the photoluminescence spectra, a top-mounted OSRAM Mercury Short Arc lamp (HBO) 100 W was properly energy filtered to provide a suitable excitation ($\lambda_{exc}$=535 nm) to excite the samples across the microscope objective. The fluorescence signal collected back through the same objective was fed into the spectrometer via a dichroic mirror (570 nm) and a long-pass filter (590 nm) to suppress residual excitation stray light. Integration time ranged from 500 ms to 2000 ms. Within the same experimental session, it was possible to record a bright-field micro image, followed by a fluorescence image and the local micro-fluorescence spectrum.

Fluorescence images (collected by means of a Nikon Digital Sight DS-2M camera) were recorded by using various excitation filters, dichroic mirrors and long-pass exit filters.

*Time-Resolved Fluorescence micro-spectroscopy:* The same microscope was used to perform pulsed micro-fluorescence experiments by exciting the sample by means of a $\lambda$=402 nm, 70 ps FWHM, pulsed solid-state laser diode head (Picoquant) at a repetition rate of 4÷5 MHz. The excitation light was fed via a monomode Thorlab optical fibre into the microscope and brought to the sample across an epifluorescence 100x objective. The fluorescence signal from the $MoS_2$ crystals was then collected back across the same objective and brought via a custom optic coupled to a Thorlabs UV/Vis fibre into a f/4, 300mm, Acton Research Spectra Pro SP-2300i triple-turret monochromator equipped with a Hamamatsu H7422-20 Peltier-cooled photosensor module, component of an inverted start-stop detection system (Picoquant, TimeHarp-100) with a 36 ps/channel time resolution and synchronously driven by an Agilent pulse generator. Transient fluorescence spectra were then reconstructed from the fluorescence transients (time range 140 ns) as collected at various wavelengths, while scanning the monochromator position from 550 to 750 nm, in 200 steps, $\Delta\lambda$=1 nm, integration time 3 s/point, using a 150 gr/mm 500 nm blaze diffraction grating. Fluorescence transients used for extracting the decay parameters were recorded for 600 s at a wavelength ($\lambda_{max}$ = 650 nm) corresponding to the maximum intensity of the emission spectrum.

# Supplementary experimental data

| *Component* | ***Pristine*** | ***EFLAO treated*** |
|---|---|---|
| $MoS_2$ | 88.6 % | 88.9 % |
| Defects | 5.4 % | 5.8 % |
| $MoO_x$ | 6.0 % | 5.3 % |
| *Ratio* | | |

| S/Mo | 1.8 | 1.9 |
|---|---|---|

**Table 1:** Relative percentage (%) of $MoS_2$ species in the pristine and EFLAO-treated samples, and semiquantitative elemental analysis extracted from spectra reported in Figure S3.

$$H_2O \rightarrow 2H^+ + 2e^- + \frac{1}{2}O_{2(g)}\uparrow \qquad \text{Water oxidation}$$

$$S\text{–}Mo\text{–}S^* + 2H^+ + 2e^- \rightarrow S\text{–}Mo\text{–}V_s + H_2S_{(g)}\uparrow \qquad \text{Desulfurization } (V_s = \text{sulfur vacancy})$$

**Scheme S1**. Proposed chemical desulphurization triggered by water oxidation.

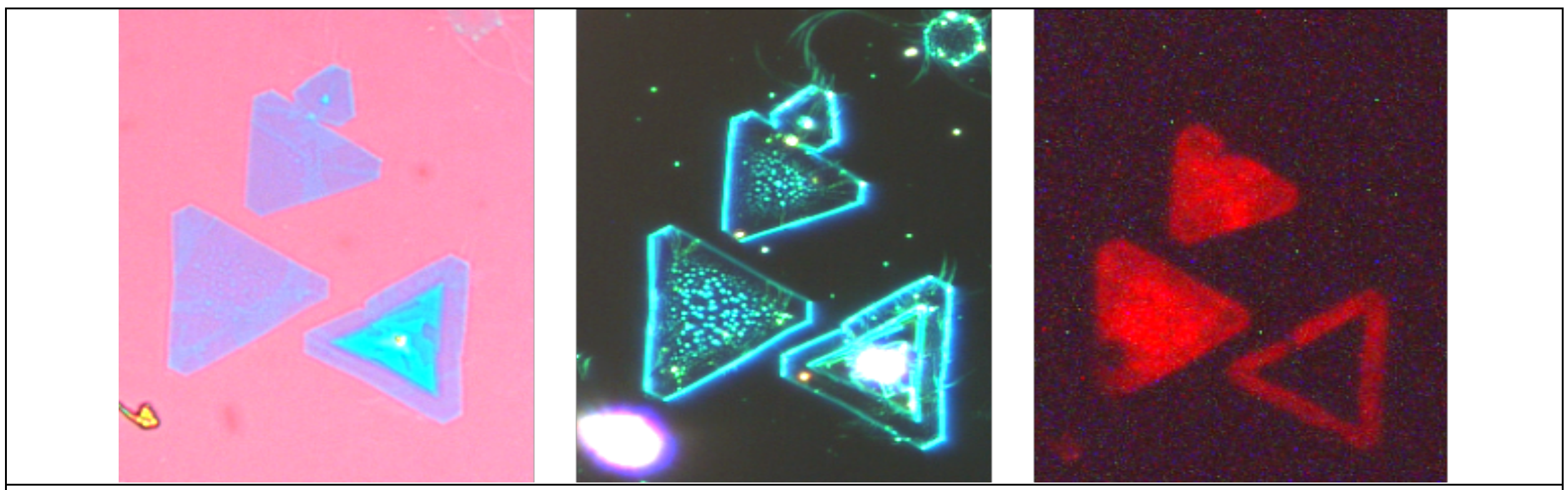

**Figure S1.** Optical characterization of $MoS_2$ monolayer crystals. (a) Optical image recorded in the bright field of $MoS_2$ flakes deposited on thermal $SiO_2$. The image shows a particular zone containing a flake with multilayers in the center and small triangular crystallites on top of the flakes (poorly visible in bright field OM). (b) Corresponding optical image recorded in dark-field. The image highlights the multi-layered structure and clearly evidence small triangular layers. (c) Corresponding fluorescence image showing the fluorescence of flake and the quenching effects of multiLayer structures and flakes.

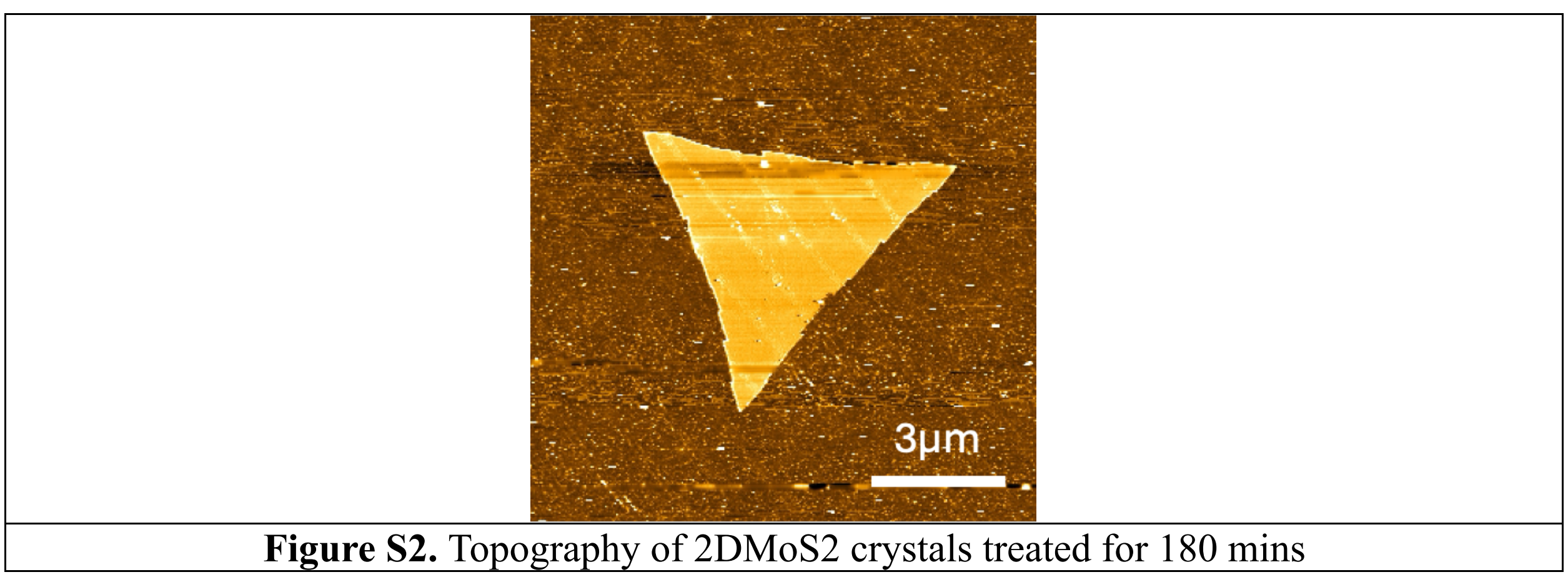

**Figure S2.** Topography of 2DMoS2 crystals treated for 180 mins

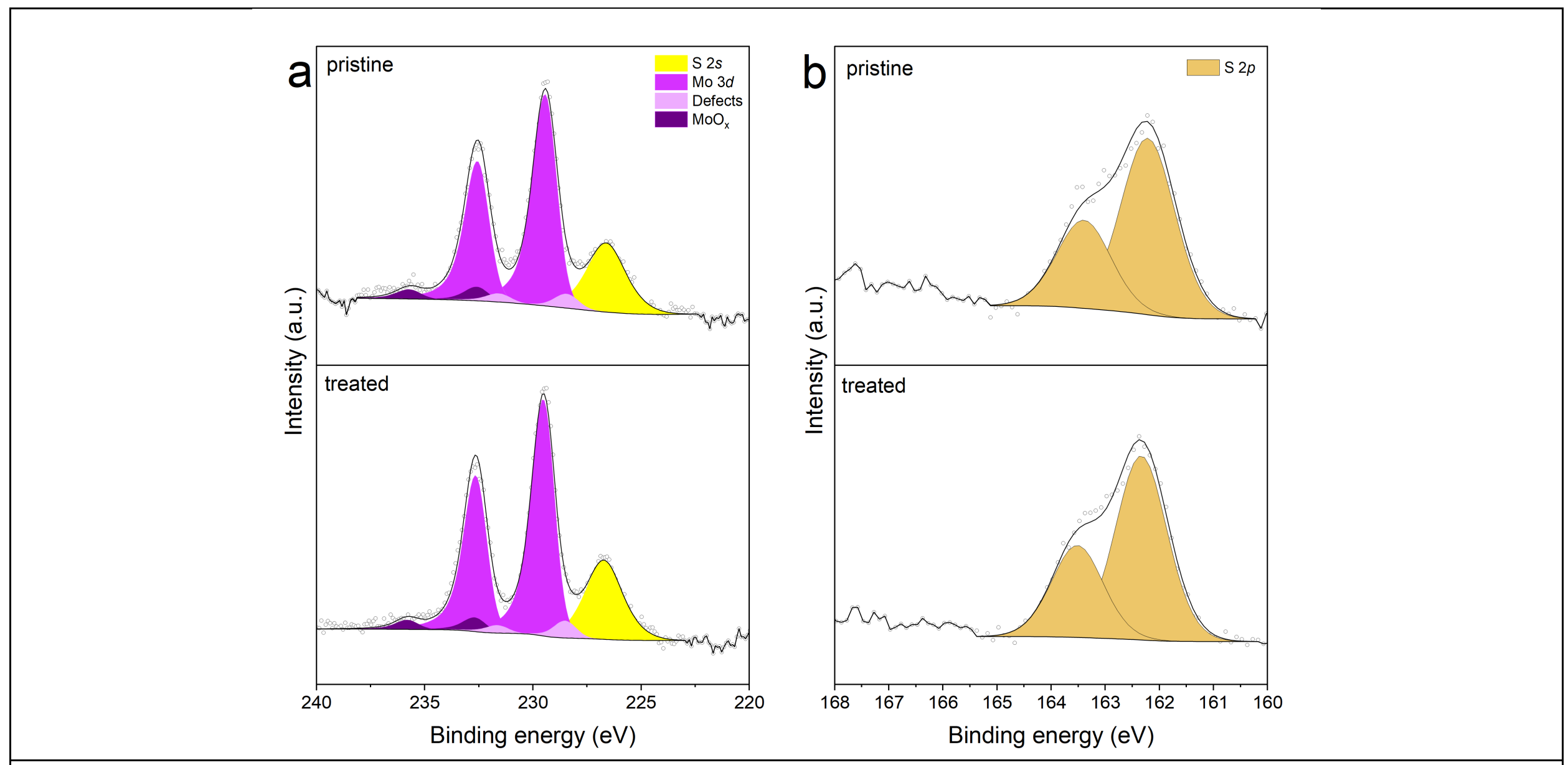

**Figure S3.** (**A**) XPS spectra of the Mo 3*d,* S 2*s* (**A**) and S *2p* (**B**) core level region prior and after the EFLAO treatment.

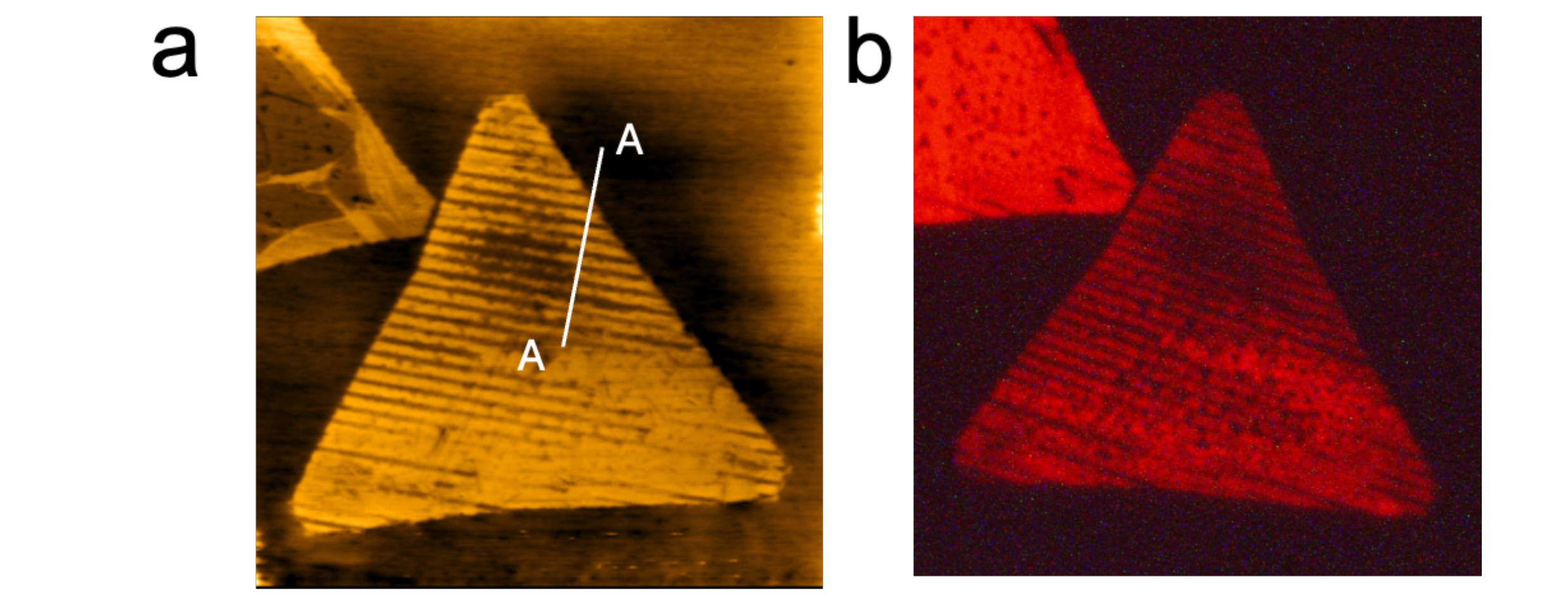

**Figure S4.** Direct comparison from surface potential images (a) and fluorescence microscopy image (b).

.

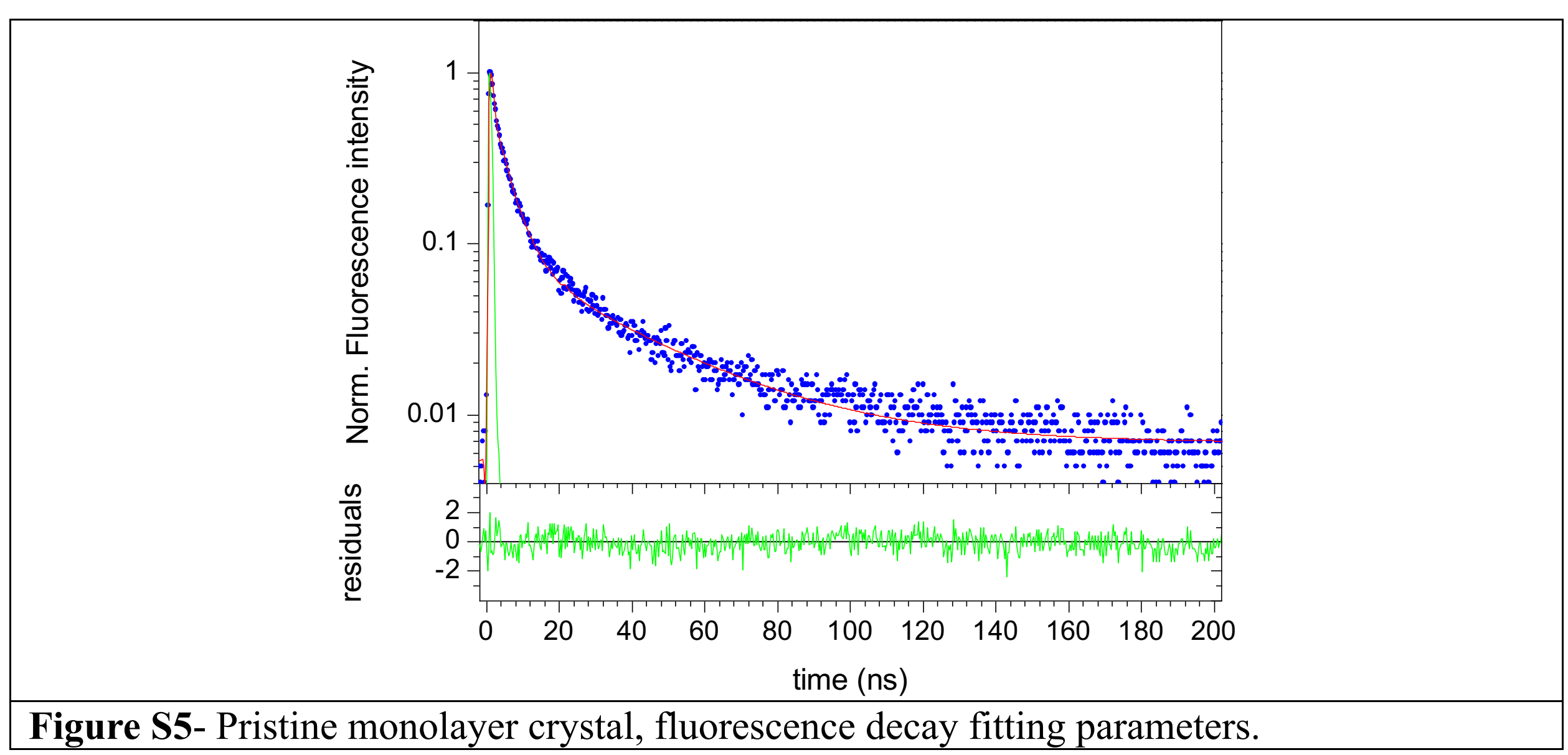

**Figure S5**- Pristine monolayer crystal, fluorescence decay fitting parameters.

**Fitting parameters:**

| | | | |
|---|---|---|---|
| $A_1$ | = | 633.4 | Cnts |
| $t_1$ | = | 4.308 | ns |
| $A_2$ | = | 2197.3 | Cnts |
| $t_2$ | = | 0.448 | ns |
| $A_3$ | = | 85 | Cnts |
| $t_3$ | = | 32.38 | ns |
| Bkg.Dec | = | 3.7 | Cnts |
| Bkg.IRF | = | -0.5 | Cnts |
| ShiftIRF | = | -86.635 | ns |

**Average Lifetime:**

$\tau_{av.1}$ = 15.669 ns (intensity weighted)
$\tau_{av.2}$ = 2.217 ns (amplitude weighted)

**Fractional Intensities of the Positive Decay Components:**

| | | |
|---|---|---|
| $t_1$ | (4.308 ns): | 42.21% |
| $t_2$ | (0.448 ns): | 15.23% |
| $t_3$ | (32.380 ns): | 42.56% |

**Fractional Amplitudes of the Positive Decay Components:**

| | | |
|---|---|---|
| $t_1$ | (4.308 ns): | 21.72% |
| $t_2$ | (0.448 ns): | 75.36% |
| $t_3$ | (32.380 ns): | 2.91% |

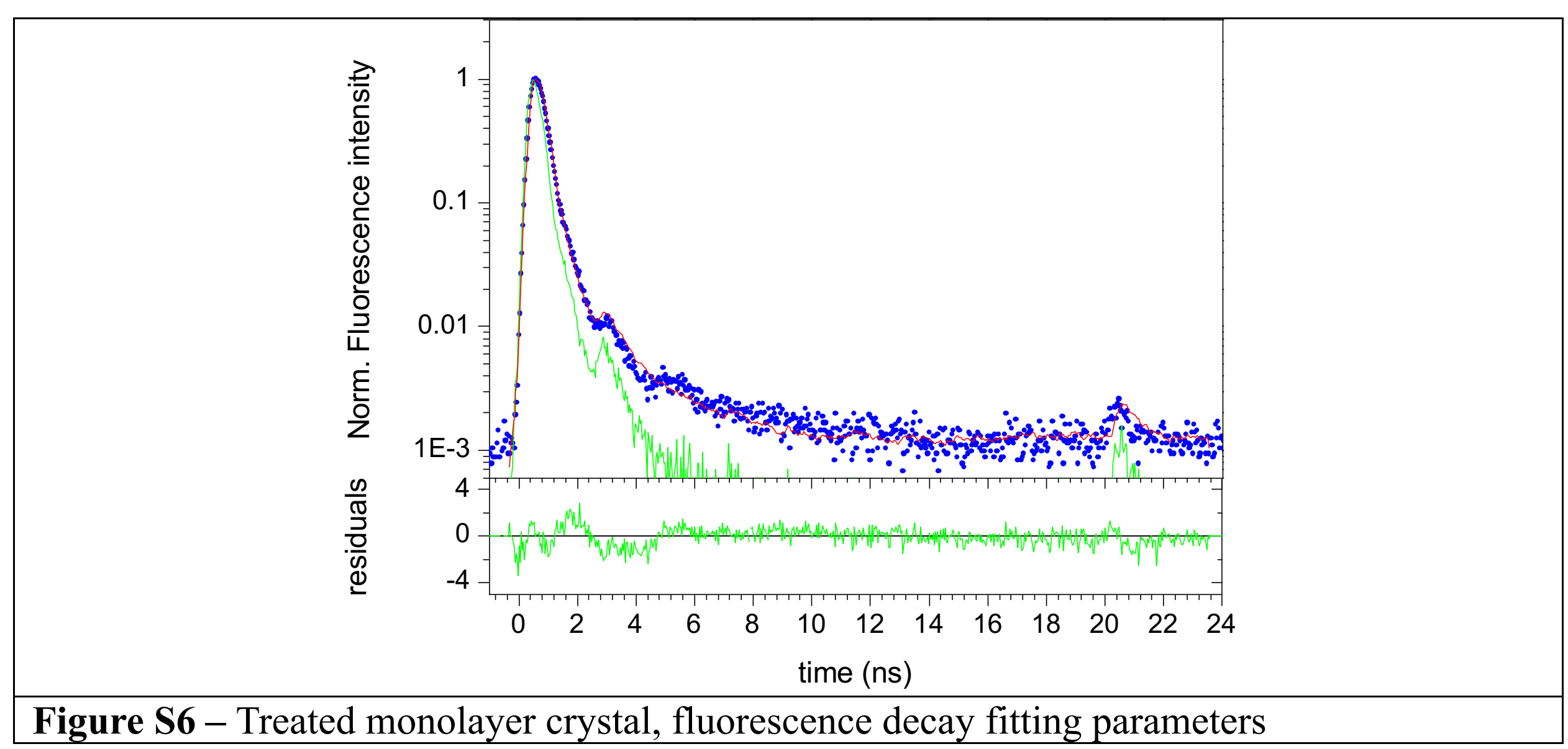

**Figure S6** – Treated monolayer crystal, fluorescence decay fitting parameters

**Fitting parameters:**

| | | | |
|---|---|---|---|
| $A_1$ | = | 120017.6 | Cnts |
| $t_1$ | = | 0.129 | ns |
| $A_2$ | = | 275.6 | Cnts |
| $t_2$ | = | 1.792 | ns |
| Bkg.Dec | = | 9 | Cnts |
| Bkg.IRF | = | -5.3 | Cnts |
| ShiftIRF | = | -0.043 | ns |

**Average Lifetime:**

$\tau_{av.1}$ =0.181 ns (intensity weighted)
$\tau_{av.2}$ =0.133 ns (amplitude weighted)

**Fractional Intensities of the Positive Decay Components:**

$t_1$ (0.129 ns): 96.92%
$t_2$ (1.792 ns): 3.08%

**Fractional Amplitudes of the Positive Decay Components:**

$t_1$ (0.129 ns): 99.77%
$t_2$ (1.792 ns): 0.23%

**References SI**